# Quantum Mechanics of Consciousness

Rajat Kumar Pradhan[1]

Vikram Dev College, Jeypore, Orissa, India-764001.

## Abstract

A phenomenological approach using the states of spin-like observables is developed to understand the nature of consciousness and the totality of experience. The three states of consciousness are taken to form the triplet of eigenstates of a spin-one entity and are derived as the triplet resulting from the composition of two spins by treating the subject and the object as interacting two-state, spin-half systems with external and internal projections. The state of deep sleep is analysed in the light of this phenomenological approach and a novel understanding of the status of the individual consciousness in this state is obtained. The resulting fourth state i.e. the singlet state is interpreted to correspond to the superconscious state of intuitive experience and is justified by invoking the concept of the universal consciousness as the underlying source of all individual states of experience. It is proposed that the individual experiences result from the operations of four individualizing observables which project out the individual from the universal. The one-to-one correspondence between the individual and the universal states of experience is brought out and their identity in the fourth state is established by showing that all individualizing quantum numbers become zero in this state leaving no trace of any individuality.

---

[1]Email: rajat@iopb.res.in

# 1. Introduction

The congenital problems of quantum theory such as the interpretation of the wave function and its so-called collapse in a measurement process have long since been associated with the possible active role of consciousness in the theory itself and also in the actual measurement process[1]. This has been exemplified in the well-known paradoxes[2,3,4] of the theory and has resulted in a variety of formulations of the quantum measurement process and a host of proposals as possible interpretations[5,6,7]. The early works of Von Neumann[8] and Wigner[9], followed by those of London and Bauer[10] and also the more recent works of Stapp[11], Mould[12], Page[13] and Zeh[14] have all conceded a fundamental role to consciousness in the measurement process.

In particular, the subject-object duality is brought to focus in the Sensible Quantum Mechanics (SQM) of Page [13] which is based on three postulates: The first one regarding the perceived object, the second regarding the perceiving subject and the third regarding their interaction or the process of perception.

Similarly, Song[15] has grappled with the problem of describing self-observing consciousness using spin-like observables and has been forced to conjecture an advantage of the Heisenberg picture over the Schrödinger picture of time evolution if it is to be described quantum mechanically as per his model.

The Weak Quantum Theory (WQT) of Atmanspacher *et al* [16]developed by relaxing and generalizing certain axioms of traditional Quantum Theory, is an attempt to basically apply quantum mechanical ideas to explain certain phenomena in psychology and psychophysiology by treating the mental states as quantum states and perception as measurement. They have also encountered the possibility of the existence of a 'collective Unconscious' as a medium and as an intermediary for the occurrence of the phenomena of 'transference', 'counter-transference' and 'deputy perception' etc. in psychotherapeutic scenarios.

A very recent and probably the most valiant of them all, is the attempt by Manousakis[17] to found quantum theory on the basis of consciousness wherein the state vector |Ψ> represents a state of potential consciousness pregnant with all possibilities, on which consciousness operates by means of a linear operator to create or modify the likelihoods of future events and thus leads to the rising of (perception of) the event in the individual observer's consciousness by comparison with the original state. The objective Universe (space-time, quantum fields including the big-bang itself) is postulated to be primarily the content of the Universal consciousness and it only secondarily gets actualized or operationally projected by consciousness itself in the central nervous system upon observation by an individual conscious observer which is called the process of perception or objectivation. The individual consciousness is taken to be a particular stream or a sub-stream of the Universal Consciousness. So, contrary to the persistent attempts to understand consciousness or its operations quantum mechanically, here quantum theory emerges as a natural description of conscious experience starting from the primary ontological character of consciousness and some of its elementary contents like perception of periodic change and motion.

If consciousness is to be described quantum mechanically, we must first of all understand what it is; what its states are; and then exploit any parallelism that obtains between consciousness and the parallel quantum systems as envisaged by Pauli and Jung [18] to analogically (rather than analytically) build up a description since the system is not amenable to sensory perception nor can it be probed by any traditional measurement, whether classical or quantum.

For our purposes, we may define consciousness as *that entity which knows or experiences*. It not only *inherently* knows itself (i.e. the subject) but also knows what is other than itself (i.e. the object) through some processes. As regards the states of consciousness we may take them to be the Conscious, the Subconscious (this includes Freud's preconscious and also the further deeper and remoter layers of retrievable memory right upto the Unconscious) and the Unconscious. The same classification, but with slightly different connotations may also be obtained from a very early and prehistoric *vedantic* text, the *Mandukya Upanishad*[S.K.19] quoted by Schrödinger[20] in the epilogue to his masterpiece '*What is Life*'. This *upanishad* does speak of the above three states of consciousness as the Waking state, the Dream state and the Deep sleep state respectively, but adds an all-important fourth state, which, in modern terminology, we may call as the Superconscious, wherein the individual becomes one with the Universal.

A phenomenological approach is introduced in this work to understand these states of consciousness in quantum mechanical terms and to derive them following the analogy with spin-like states.

After a brief introduction to the states of experience in section-2, we try to describe consciousness as a bosonic (spin-one) entity in section-3, and then move on to derive the states of experience from an interacting-fermion model of subject-object duality in section-4. In section-5, a set of four individualizing observables are introduced and their eigenvalue spectra are discussed. In section-6, the one-to-one correspondence between the states of the individual consciousness and the Universal consciousness is pointed out and in section-7, the fourth state is interpreted as a state of Superconscious experience by exploiting the identicality of the individual with the Universal consciousness when all individualizing observables vanish and by appealing to the EPR-type entanglement between the subject and the object. Finally, we conclude in section-8 with a discussion of the main results and the future direction of consciousness studies using our quantum mechanical approach.

## 2. The states of consciousness

The three distinct states or aspects of consciousness gone through regularly by each individual are the waking (conscious) state with external awareness, the Dream (Subconscious) state with internal awareness and the Deep sleep (Unconscious) state characterized by non-awareness. Before we discuss each of these states briefly to motivate the use of spin one eigenstates for their description let us note that in matters pertaining to consciousness one must have the openness to accept one's own self as the laboratory and to experiment upon oneself in a most unprejudiced manner so as to get at the truths underlying the phenomenon. Moreover, a completely objective approach to consciousness is not going to be very rewarding since it is itself the conscious subject that has developed the scientific and objective approach to understand what is other than itself. To understand consciousness is therefore the same as understanding oneself and this does not require any external aids like experimental probes. Therefore it is a foregone conclusion that objective approach to consciousness therefore is going to an incomplete affair. In any case, we begin the discussion of the states of consciousness with the hope that a quantum mechanical approach is going to help us in reconciling the subject and the object with consciousness as the underlying fundamental entity.

(i)*The (Conscious) Waking State* : In this state the *consciousness is externally projected* and there is perception resulting from *attention being fully focused upon sensory inputs* into the central nervous system in the brain. Logical ordering of events in this state leads to causal connection between a previous event and its effects afterwards. Objective space-time (i.e. separation and periodicity) along with names, forms, textures, colours, flavours and odours etc. are the contents or the *felt qualia* of the perceptions. The individual free will is most strongly felt in this state.

(ii)*The (Subconscious) Dream State* : In this state the consciousness is internally projected and there is perception resulting from *attention being fully focused upon the memory states or thought forms in the brain* which have been formed as the neural records of the previous experiences. There is *no strict causal ordering of events* as the attention shifts erratically from one memory state to another. Subjective space-time with great deal of elasticity, mental objects with adequate flexibility of form and other *felt qualia* are the contents of dream experience. The individual free will is less fully operative in the sense that we can't ordinarily direct the course of events in the dream.

According to Freud's interpretation[21], the unfulfilled desires of the waking state are sought to be fulfilled through their realisation in the dream experiences. But there can be other neurophysiological reasons for the dreams also. Further, it is generally accepted (see ref. 23, however) now that dreams are experienced during the transition from waking to deep sleep and vice versa in which there is rapid movement of the eyeballs and is therefore called the REM (Rapid Eye Movement) phase of the sleep. Modern approaches towards understanding dreams are the neurophysiological approach of Hobson[22], the neuropsychoanlytical approach of Solms[23] and the neurocognitive approach proposed by Domhoff[24].

(iii) *The (Unconscious) Deep Sleep State* : In this state the consciousness seems to be neither externally projected nor internally projected as one is completely unaware of either the external objective world through sensory inputs or the internal subjective world of impressions or thought forms recorded in the memory states. The attention seems to have lost its existence all together along with the will.

It is a state of complete ignorance of one's own self as well as of any other, as if one's consciousness is fully covered up by a thick blanket of darkness or ignorance, but is surprisingly characterized by an experience of bliss and recuperation for the fatigued individual. This state is therefore very aptly called as the Unconscious state and we have very little scope of knowing anything more than what has been stated about the contents of the experience in this state. Thus space-time and all the objects of the other two states along with their felt qualia seem to have been completely lost in the thick cover of ignorance.

It is worth noting that Freudian psychotherapy is based on the premise that the Unconscious contains many hidden data about the past experiences of individual and that the royal road to it is through the Dream or the subconscious. According to Jung[25] there is a collective or racial Unconsciousness for every species and he used this hypothesis to explain the similarities in the patterns of cultural evolution of civilizations (through the analysis of their symbols, legends, rituals and languages etc.) the globe over through millennia.

Though waking, dreaming and Sleeping are the basic states of consciousness infinitely many combinations of these states are also experienced. For example, the states of distracted attention or absentmindedness, the confused, the bewildered and the dumbfounded states, the state of obsessive thoughts, the drugged or inebriated state, the hypnotized state, states of altered perception resulting from various reasons may all be treated as having admixtures of the dream state with the waking state with various amplitudes appropriate to their experience. This is because of the fact that although we usually associate these states with the waking state, as per our description in terms of the basic states of consciousness, the awareness or attention in all these cases is only partly upon the concurrent sensory inputs and is partly on the mental impressions stored as memory. Similarly, the state of the somnambulist, the drowsy state, the lightly anaesthetized state and the state of consciousness during an epileptic fit can all be taken to be superpositions of all the three basic states with appropriate amplitudes for their experience. Similarly, the deeply anaesthetized state, the state of coma, swoon and the like can all be represented as superpositions of predominantly the Deep Sleep state with small admixtures of the Dream and/or the waking states.

It is also not an uncommon experience to have prolonged dwelling in a basic state because of our intentional or forced absorption in it, almost paralleling the quantum zeno effect-like situation with continuous observation. Similarly, very often, due to various causes we may have oscillations between waking and dream, between waking and deep sleep and so on. This is apart from the natural cyclicity of our rhythmical daily passage through these states in a fixed periodic manner. The natural cycle is from waking through dream to deep sleep and again from deep sleep back to waking through the dream state. It is to be noted that not all dreams are remembered. Only those that are immediately followed by a recollection in waking state are remembered.

## 3. Consciousness as 'light'

All the above characteristics of the three basic states of consciousness may be taken to represent the three projections- namely, the 'external' or 'spin-up', the 'internal' or 'spin-down' and the 'neutral' or 'unprojected'- of a single spin-like observable called *consciousness* corresponding to a spin-one object like a photon. Thus, the 'consciousness' quantum number has the value 1 for any individual. Therefore, we make the following associations using the $|s, m\rangle$ basis:

(a) <u>The waking state</u>: $|\omega_1\rangle = |1, +1\rangle$: The consciousness is fully externally projected leading to perception of the gross external objects through the operation of senses. Perceptions in this state are granted an objective reality in the sense of their being in existence '*even when no one is looking*' because of the sharing of the same perceptions by all waking observers concerned, though they all may not agree in regard to the felt qualia or in the details.

(b) <u>The Dream state</u>: $|\omega_2\rangle = |1, -1\rangle$: The consciousness is fully internally projected leading to perception of the subtle internal objects which are the impressions of the waking state experiences or thought forms through the operation of the subconscious mind. Perceptions in this state are granted a lesser reality compared to those in the waking state. The *dream objects* have a very peculiar kind of '*internal objective*' reality in the sense that they are open to perception by the '*dream subjects*' in the dream state. Although one may willfully enter the dream state, one cannot ordinarily direct the course of the dream because the will is incapacitated.

(c) <u>The Deep Sleep state</u>: $|\omega_3\rangle = |1, 0\rangle$: The consciousness seems to be neither externally projected nor internally projected leading to non-perception of either the gross external objects or the subtle internal objects. Instead, there is a covering of *blissful ignorance* upon the awareness. This seems to be a kind of unconscious state because of the non-awareness of even one's own self and looks like an unprojected state of consciousness. Still, this is not the state of '*consciousness as it is*', since consciousness, by definition, must have self-awareness.

Again, we can't say that '*consciousness as it is*' was absent during deep sleep for although individual self-awareness was lost, the experience of a blissful sleep could somehow be registered in this state and recovered also on waking. If everything was obliterated in the thick cover of ignorance or Unconsciousness how is it that the individual that wakes up afterwards is the self-same individual that went into Deep Sleep? Thus, the individual's memory states are not deleted in Deep Sleep But are kept in a kind of suspension. The Consciousness is temporarily suspended or withdrawn from both– the internal memory states and the external sensory inputs.

This is the first indication that the above three states do not fully exhaust the possible states of experience. There must be a fourth state of unprojected consciousness which is fully self-aware so that we can identify it as consciousness *per se* or pure consciousness.

The second indication is from the sequence of transitions among the states that we experience. To see this, let's assume the triplet of eigenstates to be a complete orthnonormal set and write the general state of consciousness $|\Psi\rangle$ as a linear superposition of them:

$$|\Psi\rangle = a_1 |\omega_1\rangle + a_2 |\omega_2\rangle + a_3 |\omega_3\rangle \quad \ldots \ldots \ldots (1)$$

where, the expansion coefficients $a_i$ are such that $|a_i|^2 = |\langle \omega_i|\Psi\rangle|^2$ gives the *intensity of experience* of the state $|\omega_i\rangle$, when the consciousness is in the state $|\Psi\rangle$. The orthonormality is expressed by $\langle \omega_i| \omega_j \rangle = \delta_{ij}$.

Now, the sequence of experience as delineated in the previous section is: $|\omega_1\rangle \rightarrow |\omega_2\rangle \rightarrow |\omega_3\rangle \rightarrow |\omega_2\rangle \rightarrow |\omega_1\rangle$ i.e. $|1,+1\rangle \rightarrow |1,-1\rangle \rightarrow |1, 0\rangle \rightarrow |1,-1\rangle \rightarrow |1,+1\rangle$, which means that in the transition from waking to dream and vice versa the selection rule $\Delta m = 0, \pm 1$ for $\Delta s = 0$ is violated, assuming that they are like the well-known radiative transitions. But, the fact that we experience these transitions tells us that they are not forbidden and hence, there must be a state with projection zero which intermediates these transitions and is also available as an alternative route for each allowed transition. This state can only be the $|0, 0\rangle$ singlet state, discarding the possibility of $|2, 0\rangle$ which would lead to an unnecessary proliferation of states contrary to experience.

Incontrovertible experience of the individual during the transition from waking to dream testifies to the above fact because as one gradually withdraws oneself into the dream state, one cannot do it keeping up a continuity of thoughts or a continuous movement through the space of memory states since there is a momentary loss of individual consciousness for a fleeting moment– a momentary blackout, so to say– just before the rise of the dream consciousness. This is to be experienced by waking up in the middle of the dream, before one lapses into deep sleep. Similar is also the case during the inverse transition i.e. we do not come to the waking state keeping up a continuity of thought flow or experience from the dream state. On the contrary, if it were possible for the individual to keep up the chain of thoughts right into the dream state one would always succeed in dreaming exactly to one's liking! Similarly if it were possible to actuate the dream experience by continuity into the waking we would get the most pleasant dream experiences actualized on waking up by keeping up such continuity! But, this is not the case.

This clearly shows the distinction between the waking and the dream states in addition to bringing home the necessity of a fourth state. The dwelling in the fourth state during such transitions is so ephemeral that it passes of unnoticed and does not interfere much with remembrance of the dream upon waking up. The same fourth state is also gone through in a flash during the transition from one thought form to another so quickly that it is never suspected to have been there at all. What exactly is the experience in such a flash? Because the dwell time is extremely short we have no way of answering this question unless we somehow master the practical technique of prolonged dwelling in it. We shall, however, attempt to provide a theoretical framework for understanding this state in section-7.

Yet another reason to seek for the fourth state is that if these three were the only possible states, then we have the same problem of quantum jumps as in old quantum theory since we have no answer to the question as to where the consciousness lies during a transition from one thought form to another thought form in the waking and the dream states and also during a transition from one state to another. In quantum field theory, however, the existence of the 'vacuum state' comes to the rescue because we interpret the transition from state |1> to state |2> as annihilation of the system in state |1> and its subsequent creation in state |2> through the annihilation and the creation operators, so that we can safely say that during the said interval, the system temporarily merges into the vacuum before emerging again back into existence in the new state |2>. Thus, we need to have the so-called 'vacuum' state which will serve the purpose of being the source and the substratum for the three states that we normally experience.

So, these observations on a possible fourth state lead us to try to understand the *whole of experience* in an interacting-fermion model based on the subject- object duality which happens to be fundamental to all experience.

**4. Spin-half realisations of the subject-object duality**

The experiences of the individual subject in the states $|\omega_1>$, $|\omega_2>$ and $|\omega_3>$ are characterized by the facts of the individual's awareness of himself or self-awareness and of what he considers as 'other than himself' or the object. The subject and the object make up the whole of our individual experiences. In $|\omega_1>$ and $|\omega_2>$, both these awarenesses are present, but in $|\omega_3>$, although neither of them is manifestly present, somehow the individual's experience of bliss is recorded. In all the three states, we may consider the subject and the object to have independent existence with corresponding **'existence'** quantum numbers $e_s$ and $e_o$ taking values half each ( i.e. $e_s = ½$ and $e_o = ½$), since *existence is the common characteristic of both*. The subject and the object are distinguishable only on the introduction of another observable (a new quantum number) **'consciousness of existence'** having values 1 and 0 respectively for them.

Now, the individual subject and the object (neither of which is experienced in $|\omega_3>$) are experienced in their two possible projections corresponding to 'existence-up' with external projection $+½$ in $|\omega_1>$ and 'existence-down' with internal projection $-½$ in $|\omega_2>$. The most general subjective state can be represented by

$$|\Phi> = c_+ |½, +½> + c_- |½, -½> \qquad \ldots \ldots \ldots (2)$$

and, the most general objective state by:

$$|\chi> = d_+ |½, +½> + d_- |½, -½> \qquad \ldots \ldots \ldots (3)$$

where, $c_\pm$ and $d_\pm$ are the '***existence amplitudes***' in the respective subjective and objective states of external(+) and internal(−) projection.

In $|\omega_1>$, the externally projected individual subjective existence experiences the externally projected objective existence, while in $|\omega_2>$, the internally projected individual subjective existence experiences the internally projected objective existence. In the product basis, these two states can be represented by $|m_s, m_o> = |+½, +½> = |e_s = ½, m_s = +½>|e_o = ½, m_o = +½>$ and $|m_s, m_o> = |-½, -½> = |e_s = ½, m_s = -½>|e_o = ½, m_o = -½>$. But, experience is impossible unless there is some kind of interaction between the individual subject and the object, both of which we have assumed to have independent existences.

The simplest kind of interaction is of the familiar **L·S** –type, which, in this case we put as:

$$V_i = K_i \, \mathbf{e}_s \cdot \mathbf{e}_o \qquad \ldots \ldots \ldots (4)$$

where, $K_i$ is a coupling parameter that may have factors depending on the space, time and other variables of the individual's waking state($x$, $t$; $q$); space, time and other variables of the dream state ($x'$, $t'$; $q'$), and the variables characterizing the experiences in Deep Sleep(see section-5).

Following the well-known procedure for the composition of angular momenta, we can now switch over to the total angular momentum basis or $|j, m\rangle$ basis where the interaction will be diagonal. We name this basis as the **'Experience Basis'** and write the resulting orthonormal eigenstates viz. the triplet (corresponding to existence-consciousness value 1) and the singlet (corresponding to existence-consciousness value 0) eigenstates of individual experience as follows:

$|\omega_1\rangle = |1, +1\rangle = |+½, +½\rangle$

$|\omega_2\rangle = |1, -1\rangle = |-½, -½\rangle$

$|\omega_3\rangle = |1, 0\rangle = (1/\sqrt{2}) \{|+½, -½\rangle + |-½, +½\rangle\}$

$|\omega_4\rangle = |0, 0\rangle = (1/\sqrt{2}) \{|+½, -½\rangle - |-½, +½\rangle\}$

where, we have identified the triplet of symmetric eigenstates with the consciousness eigenstates discussed earlier. The antisymmetric singlet, we have identified as the fourth one, the reasons for the existence of which were also discussed in section-3.

How far are we justified in making these identifications? While there is no problem with the first two identifications, we do have to see what new understanding of the deep sleep state this analysis grants us, which is one of the reasons for applying the interacting-fermion model for the subject-object duality. We also need to see whether the fourth state really serves its purpose as discussed earlier and what its implications are for understanding the link between the individual consciousness and the Universal consciousness (See section-7).

To understand the experience of Deep Sleep in this model we see that in $|\omega_3>$, there is symmetry between the subject and the object in regard to the interchange of their projections. This means that we can interpret the Deep Sleep experience as one in which the externally projected subject ($e_s = ½$, $m_s = + ½$) does not have any externally projected object ($e_o = ½$, $m_o = + ½$) which it could have experienced; instead it has only the internally projected object (eo = ½, mo = −½) available to it. Similarly, the internally projected subject ($e_s = ½$, $m_s = - ½$) does not have any internally projected object ($e_o = ½$, $m_o = - ½$) which it could have experienced; instead it has only the externally projected object ($e_o = ½$, $m_o = +½$) available to it. Thus, a state of non-experience or non-perception results for the individual consciousness.

As an interesting aside, we may consider the question as to how one wakes up to $|\omega_1>$ from $|\omega_3>$ on being called by name or on being given some other input in general, sufficient for the purpose, if the above explanation for non-perception in $|\omega_3>$ is assumed to be correct. The answer lies in the fact that any (sensory, vital or mental) input strong enough to warrant a premature or forced or induced transition from $|\omega_3>$ has to be through the intermediate Dream state $|\omega_2>$ which means that it has to come via internal perception through internal projection. The time spent in $|\omega_2>$ in this case may be a very tiny fraction of a second before it takes cognizance of the strong external sensory input or the internal (vital or mental) inputs. In essence, what happens is that the internalized part of the subject is provided with some internal object and the externalized part is provided with some external object, thus making perception in $|\omega_1>$ through $|\omega_2>$ possible.

We may also picture the time evolution by using time-dependent coefficients $a_i(t)$ multiplying $|\omega_i>$ in an evolving superposition

$$|\phi(t)> = \sum_{i=1}^{3} a_i(t) |\omega_i>  \quad \ldots \ldots \ldots (5)$$

during the intermediate state such that $|\phi(t_0)> = |\omega_3>$ and $|\phi(t_1)> = |\omega_1>$. This is how one wakes up (see however, section-7 for the explanation basing on the alternative path through $|\omega_4>$).

This is quite a novel understanding of the so-called unconsciousness of Deep Sleep that emerges from our modelling of individual experience as above. The zero projection is reflective of the failure of the subject to make contact with the object and as a result, it fails to know the object, which in turn leads to lack of self-awareness of the subject since the self-awareness of the subject is dependent on its awareness of the corresponding objects. This is because the self-awareness experienced by the subject in waking and dream is always in conjunction with the awareness of the corresponding waking or dream object as evidenced by the experience eigenstates |$\omega_1$> and |$\omega_2$> listed above. The awareness of oneself is always inextricably associated with the awareness of 'what is other than oneself' and in |$\omega_3$>, both these are absent. This is the explanation of the state of non-awareness of Deep Sleep.

## 5. The Individualizing observables

It is a well-known fact that all individuals do not have the same objective experiences although all may agree on certain aspects of the objective reality. We do not all agree fully with each other on the *felt qualia* that we associate with the objects. At the same time, it cannot be gainsaid that we agree on certain very important characteristics of objects and this partial unanimity is what is at behind our granting an objective reality to them independent of individuals. If we assume that there is indeed an objective reality independent of individuals, as we do in the scientific approach to reality, then naturally, we must ask, 'what causes the differences in the individual perceptions of the same objective reality?' Obviously, there must be some differing characteristics in the individuals themselves which are responsible for the differences in their perceptions. What will be quantum mechanical description of such individualizing characteristics that lead to the multitude of individuals?

In what follows, we propose to explain the multiplicity of individual perceptions by adopting a set of four mutually commuting ***individualizing observables*** **A**, **B**, **C** and **D**; (hence, we shall call them the '**ABCD- observables**') which have different values for different individuals. Their eigenvalues are the characteristic 'quantum numbers' of the individual exactly like the mass, charge, spin and other quantum numbers associated with quanta. These may be taken to be:

***(a) Attitude***: The operator **A** represents the attitude of the individual towards what it considers as '*the* other' and therefore, may be one of attraction (love) which we shall take to be a 'positive' attitude (a>0) and aversion (hate) which we shall take to be a 'negative' attitude (a<0) and finally, indifference or neutrality which we shall take to be the zero attitude (a=0). '*The other*' referred to above may be any *spatiotemporally limited expression of the Universal Being* i.e. it may be a felt quality (a sound, texture, color, flavor or odor or a virtue or a vice), an event or a process, or a living or nonliving entity or group of such entities or any experience in general. To be explicit, we may test the attitude of an individual towards (a) ethical living by sincere performance of duties (b) acquiring wealth through righteous means and (c) fulfilling the vital and emotional urges within limits. These may be termed as the *ethical*, the *economic* and the *emotional* attitudes respectively. As noted above, these attitudes may be positive or negative or neutral. Our attitudes towards all objects, qualities, events, processes and experiences will be contained as special cases of these three basic attitudes. Thus, unethical living, greed for amassing wealth through unrighteous means and unbridled sensual gratification will correspond to negative attitude values.

Broadly speaking, the attitude can be figured out by eliciting responses from an individual to the questions as to whether there is a *most liked* and a *most disliked* 'something' through any method (observation, questionnaire, schedule, interview or from primary or secondary sources, or any combination of them, as the case demands ) appropriate for the purpose[26], and then points may be awarded depending on whether the individual finds any least likable and least dislikable characteristics (or felt qualia) in the most liked and the most disliked respectively. Thus, we can represent all individual attitudes in the range [−1, +1], with the extrema ±1 corresponding to unqualified infatuation and unqualified aversion respectively. Further, in our approach, it is immaterial what object or person or quality or event or process or experience one likes or dislikes most but that there are such *most liked* and *most disliked* 'something' is important for fixing the values of the attitude.

Generalities apart, the experiences of the individual are in accordance with his likes and dislikes and this is what creates the special differences in his/her perceptions compared to others. Moreover, it is to be kept in view that while the strength or intensity of the individuality is to be judged from how strong the likes and dislikes are, the mere non-vanishing of the Attitude (or any of the individualising observables in general) howsoever infinitesimally close to zero it may be, is sufficient for the projection of the individual from the Universal.

This is to be contrasted with what in modern psychology is termed as the 'Diamond of Opposites'– a method of determination of attitudes by plotting the attraction and aversion along orthogonal axes to form a diamond[27]. However, our concern here is only with whether a person has any likes and dislikes or not, and if yes, how intense are the strongest attraction and the strongest aversion, no matter towards what object, quality, event or process or experience such attraction or aversion is directed. The greater the ignorance, the stronger is the distinctive ego and accordingly the stronger are the likes and dislikes.

The attitude determines most of our conscious and subconscious activities in the waking and dream states respectively. These actions then lead to further accentuation or strengthening of the likes and the dislikes. Acting as per these strong likes and dislikes becomes our habit and our habits go to form our character which shapes up our future evolution or destiny. Obviously then, the zero eigenvalue corresponds to a special kind of individuality, which is without any attractions or aversions and thus, may well be taken to be equivalent to the state of Universality since the Universal being all-inclusive has no '*other*' to either love or hate.

We note that we can always split the positive and the negative halves and get two observables **A₊** and **A₋** corresponding to attraction and aversion respectively so that the range of each will be restricted to the interval [0, 1]. This may be done to have completely identical eigenvalue spectrum [0, 1] for all the five individualizing observables namely, **A₊**, **A₋**, **B**, **C** and **D**.

***(b) Body-identity***: Almost all individuals are characterized by their complete identification with the bodily personality. Rare exceptions occur only in very special situations (e.g. in the mother for the protection of her child, in the soldier for the protection of the territorial integrity of the country, in the friend for the wellbeing of the friend etc.); or in very exalted selfless individuals (like Jesus Christ), who may happily undergo bodily suffering for any good cause. Each of us knows how dear the body is to us and how very '*exact*' is our identification with it.

Thus, we may take the observable **B** to have eigenvalues in the range [0, 1], the eigenvalue zero again being a very special occurrence, almost coinciding with Universality. Obviously, the density of states (individuals) will be very high near the eigenvalue b=1. It may be noted here that our identification in waking and dream is with our own gross (or physical) and subtle (or mental) bodies respectively, while in deep sleep, we are one with our own ignorance which is the very cause of our individuality (hence named as Causal Ignorance).

*(c) Causal Ignorance*: This observable has the eigenvalue 1 for the state of deep sleep and a value less than one in waking and dream. The eigenvalue zero is again a very special one corresponding to complete removal of all ignorance and therefore, to complete knowledge or omniscience! The spectrum of eigenvalues for the observable **C** is thus the interval [0, 1]. The state of complete knowledge (c = 0) must be one devoid of any individuality since the individual is always characterized by limited knowledge because of its dependence on the senses, the mind and the intellect etc. and their various modes (space, time and causation etc.) for acquiring knowledge. The individual is further handicapped by its point-like location and the inability to perceive beyond certain allowed ranges of vibratory inputs through the various senses or through the measuring instruments which are only the 'extended senses'.

On the contrary, the Universal is everywhere present and has all-knowledge through simultaneous direct contact or perception of all causes and effects spread over the entire spacetime domain. No wonder that this seems impossible for the individual to visualize because it is simply not meant for individuals like us to visualize. Just as one species cannot visualise the perceptions of another because of the lack of the appropriate organs, similarly also we individuals cannot visualise the workings of the Universal. In fact, even an advanced and far more evolved being like the human being fails to visualise the perceptions of elementary living entities like small bacteria or an ant. What to speak then of the visualization of the higher's perception by the lower, or more so, of the Universal by the Individual?

This Causal Ignorance (**C**) is itself of the form of bliss– bliss of one's being a separate individual entity characterized by the Distinctive ego (**D**) and the bliss of having this individuality manifest through the Attitudes (**A₊** and **A₋**) and the resulting Body-identity(**B**) through which one associates oneself with or experiences this bliss. This Causal ignorance **C** is therefore the most fundamental of these individualizing observables, since it is, in the sense just described, the cause of the rest of the observables and consequently of all experience. It is the deepest reason for the appearance of the individuality and is the cause of the Distinctive Ego **D**. Thus the value c=1 corresponds also to a state of the highest bliss in addition to being the state of the highest ignorance or non-awareness.

Since, c=0 corresponds to the Universal, the Causal ignorance may therefore, be said to be nothing but the ignorance of this state of universality which is a possibility for the experiencing individual to evolve into, by gradually reducing all individualizing quantum numbers to zero. We mostly spend our lives in the first three states, hardly ever worrying about the fourth, except in the very trying of circumstances or on very rare occasions of deep introspection on the origin of joy or grief.

*(d) Distinctive Ego*: That which separates the individual from the rest of the Universe is the sense of being a separate entity. Out of Causal Ignorance arises this kernel of one's individuality, the Distinctive Ego (or the Differentiating Ego) represented by the observable **D**. Ordinarily, this has the value 1 in $|\omega_1\rangle$ and $|\omega_2\rangle$, while in both $|\omega_3\rangle$ and $|\omega_4\rangle$ it has the value 0. Its separative effect is realized through the operations of the attitudes $A_\pm$ and the Body-identity **B**. One then considers oneself as a finite, limited individual living in a certain external spatiotemporal domain physically and having certain recorded experiences in the internal i.e. mental domain. Then follow the notions of one's personal (i.e. physical height, color, sex, age, bodily appearance etc.), familial, racial, territorial, national, earthbound *external identity* as well as the vital, mental, intellectual and the experiential *internal identity*. One then, for all practical purposes, behaves as an individual, limited by one's own identifications with finite domains of space-time and consciousness. This is where we find ourselves operating as normal individuals with fully active individualities corresponding to d $\cong$ 1.

The eigenvalue spectrum for this operator is thus the interval [0, 1]. Again, we see that when d is zero, there is complete lack of the ego sense and the individual expands out into the Universal and attains oneness with it. In Deep Sleep, because of Causal ignorance one does not know this expansion into, and the oneness with the Universal, but in $|\omega_4\rangle$ it is not so and hence we may identify it with the state of Universality of being.

Now, we need to address the question of the actual measurement of these individualizing observables, so that they qualify to be observables with some sort of 'objectivity' and exactness within allowable limits, in order to qualify for application in a scientific investigation. Obviously, these are not observables like energy or momentum so that we can use measuring instruments and get their values. Here, what we have to adopt are the well-known techniques (viz. observation, questionnaires, schedules and interview methods etc. or their combinations) employed by the researchers in the so-called 'inexact' sciences (i.e. humanities) like economics, psychology, sociology, management studies, medical science etc. We may prepare intelligently designed questionnaires appropriate to the observable concerned, with full points 100, and from the responses from an individual, we may get the value of the observable in the waking state by scaling down to the eigenvalue range [0, 1]. Interestingly enough, several methods have been devised in the field of psychology to determine the individual attitudes **$A_\pm$**, which may be taken as the guiding principles for determining the other observables as well, but keeping in view the specific requirements of the quantum mechanical formulation presented here. Starting with Thurstone's equal appearing interval scale[28, 29] to quantitatively represent attitudes, we have the summated rating scale of Likert [30], the cumulative scale of Guttman[31] and the Semantic differential technique[32, 33] using bipolar adjectives etc. in addition to the Diamond of opposites mentioned earlier, for the determination and representation of attitudes.

For the use of **$A_\pm$** as quantum mechanical observables characterizing the individuality, we need only the maxima of the attraction and aversion towards any experience in the past, present and future, sensory or otherwise.

The values of the observables in the Dream state (subconsciousness) may be inferred from the corresponding values in waking with fair amount of accuracy if we take note of the fact that the waking and the Dream states mirror each other in the sense that one's sub-conscious thoughts get manifested in the waking as conscious actions, *modulo* social or environmental restrictions.

## 6. The individual and the Universal

The preceding discussion goes to show that there is a very special kind of state corresponding to the vanishing eigenvalue for each of the individualizing observables, which we have referred to as the Universal state. The existence of a Universal consciousness has been taken as an essential ingredient in Manousakis' formulation of Quantum theory on the basis of consciousness.

However, a small but significant difference exists between the collective and the universal states although they have been used interchangeably in the literature. The collective is the sum of the individual experiences while maintaining their distinctive individualities ($d \neq 0$), the Universal is the melting pot of all individual experiences wherein all distinctive individualities vanish *in toto*. The collective is a subset of the Universal in the sense that the collective may refer to particular group like a family or a clan or a race, while the Universal always refers to the totality of all individual experiences.

Jung's insight[25] of the collective Unconscious may be seen as a limited expression of the Universal unconscious. If all Individual unconscious is comprehended in the collective or universal unconscious, then similarly also, all individual conscious and subconscious contents must be comprehended in the corresponding Universal conscious and subconscious, since, if at all the Collective Unconscious is manifest in the cultural evolution of different civilizations then it must have done so through the layers of the collective subconscious and the collective conscious which are nothing but the limitations of the Universal consciousness and subconsciousness corresponding to a species. This also paves the way for postulating a one-to-one correspondence between the individual and the Universal states of experience.

Our approach to understand experience using spin-like observables also points to the existence of the Universal state of consciousness as the fourth state shorn of all individuality.

The simplest, most straightforward yet profoundest reason for the existence of the state of Universality is as a source for all individual experiences, for otherwise, it would be impossible for us to account for the continuity and the regularity of the pattern of our daily experiences. Although the individual consciousness seems to be absent in Deep Sleep it must be there in some form lest one would not wake up as the same individual that went into Sleep. Thus, there must be continuity of the individual consciousness at a level deeper than these three levels of daily experience. Whence comes this deepest individual consciousness or the core consciousness and where does it reside and how long? How is it able to project itself unto the different states of experience? These are still deeper questions to be answered. It is elementary common sense knowledge that the higher (i.e. more expanded) forms of consciousness include and transcend the lower (less expanded) forms, in the sense that the latter are fully comprehended in the former. This gives us a clue towards the existence of the most expanded form of consciousness which includes in itself all limited manifestations of consciousness without being exhausted by them. This is the state of the Universal Consciousness or Cosmic Consciousness which includes in its bosom all states of limited expression of itself in the entire life of the Universe and yet transcends them. Therefore, for all practical purposes, it can be taken to be an infinite, inexhaustible and eternal consciousness.

In our phenomenological approach, we can build up the link between the individual and the universal by postulating a one-to-one correspondence in their states of experience. There must be the four states of Universal Consciousness $|\Omega_i>$ corresponding to the four states individual consciousness $|\omega_i>$, i= 1, 2, 3 and 4. The individual states are the experiences of the same One Universal Consciousness with the individualizing observables $A_\pm$, **B**, **C**, and **D** taking up different non-vanishing values, thereby lifting the degeneracy for different individuals. The continuous interval [0, 1] for their eigenvalues allows for the infinite of individuals to be encompassed by the Universal as required. These observables therefore serve to limit the unlimited, the infinite and eternal Universal consciousness to finite, spatiotemporally limited individuals.

The four Universal states of experience $|\Omega_i>$ may be seen to be the result of the interaction between the Universal Subject and the Universal Object each carrying Existence a value ½, exactly similar to the composition of existences for the individual subject and the corresponding object of individual experience leading to the individual states $|\omega_i>$. It is to be noted in this regard that the individual subjects and their objects are not to be summed arithmetically to get the Universal Subject and the Universal Object. Just as the existence of the manifold sates of the harmonic oscillator does not give us infinitely many such oscillators, similarly also these individual states are the expressions of the One Universal Consciousness but with different individualizing quantum numbers for different individuals. The individuals in this sense are mere limited or finite appearances of the Universal.

This is to be contrasted with Manousakis' view of the individual streams of consciousness being particular sub-streams of the Universal stream, which may give the impression that the Universal may be gotten by adding up all the individuals. Our phenomenological approach reveals that the Universal whole is not the sum of the individual parts. Rather, all the parts put together can never exhaust the Universal for the simple reason that the former are the appearances of the latter. This follows from our postulated correspondence between the individual and the Universal states of Experience which, in turn, requires the Universal Subject and the Universal Object to be represented as spin ½ existences. However, this cannot be the case if we simply algebraically add the infinite number of individual subjective and objective spin ½ existences following angular momentum addition rules.

To represent the relation of the individual with the Universal in quantum mechanical terms, we denote for the sake of brevity, the n[th] individual by the set of quantum numbers I[n] corresponding to the values of the set of individualizing observables $I^n$ = {$A_\pm^n$, $B^n$, $C^n$, $D^n$}, so that we may write the experiential state of the n[th] individual in the eigenstate |ω$_i$> as:

$$|\omega_i^n> = |\Omega_i; I^n> \qquad \ldots \ldots \ldots (6)$$

And,

$$I^n |\omega_i^n> = I^n |\omega_i^n> \qquad \ldots \ldots \ldots (7)$$

The most general time-dependent state of the individual will be given by (5), but now with the additional individuality index 'n':

$$|\phi^n(t)> = \sum_{i=1}^{3} a_i(t) |\omega_i^n> \qquad \ldots \ldots \ldots (8)$$

where, the fourth state is left out of the superposition for the reason that it is not an ordinarily experienced state for the individuals who follow their daily rounds of waking, dream and sleep and their various superpositions throughout their lives. If at all it is experienced, it is only fleetingly with lifetime $\tau_4 \ll \Delta t_0$, the observation (measurement) time (see section-7). It is to be noted that the parameter t in the equations is the waking time because all our quantum theory is done in $|\omega_1\rangle$.

Since there are a countable but very large number of individuals constituting the Universal, we may be interested in a density matrix representation of the quantum states of the latter. However, It is easy to see that because mostly the individuals spend their time in psychophysical superpositions $|\phi^n(t)\rangle$, a simple density matrix for a total of $N$ individuals like :

$$\boldsymbol{\rho}_N = \sum_{i=1}^{4} n_i |\omega_i\rangle\langle\omega_i| \qquad \ldots \ldots \ldots (9)$$

with weights $n_i = N_i/N$, $N_i$ being the number of individuals in the eigenstate $|\omega_i\rangle$, used ordinarily in quantum mechanics to describe macroscopic systems would not suffice for the description of the Universal state comprising of these individuals.

We may, instead, introduce a 'phenomenological density matrix' to build up the Universal experience from the individual experience eigenstates as follows. First of all, we note that the state $|\phi^n(t)\rangle$ is different for different individuals although the coefficients $a_i(t)$ may be the same at the time t because of the characteristic peculiarities of each individual's experiences due to the differences in the values $I^n$ of the individualizing observables. Thus, there will be as many states as are individuals in the summation and hence the weight factor for each term (individual) will be $\frac{1}{N}$. Therefore, we write the density matrix for N individuals as:

$$\boldsymbol{\rho}_N = \frac{1}{N}\sum_n |\phi^n(t)\rangle\langle\phi^n(t)| \qquad \ldots \ldots \ldots (10)$$

Or, in terms of the individual eigenstates $|\omega_i^n\rangle$:

$$\boldsymbol{\rho}_N = \frac{1}{N}\sum_n \sum_i \sum_j a_i a_j^* |\omega_i^n\rangle\langle\omega_j^n| \qquad \ldots \ldots \ldots (11)$$

Since, we are building up the Universal phenomenologically we do not necessarily have to take the N→∞ limit in the sum. For a single individual, we see that the 'phenomenological density matrix' is just the projection operator for the experiential pure state $|\phi^n(t)>$ :

$$\boldsymbol{\rho}_1 = P_n = |\phi^n(t)><\phi^n(t)| = \sum_i \sum_j a_i a_j^* |\omega_i^n><\omega_j^n| \qquad \ldots \ldots \ldots (12)$$

Further, using the orthonormality [34] of the individual states $|\phi^n(t)>$:

$$<\phi^m(t)|\phi^n(t)> = \delta^{mn} \qquad \ldots \ldots \ldots (13)$$

we can readily verify the well known properties of the density operator for pure and mixed states holding for $\boldsymbol{\rho}_1$ and $\boldsymbol{\rho}_N$ ($N \geq 2$) respectively.

Here itself, we see clearly the infinitude of the possibilities inherent in the Universal even for one experiencing individual since as the operator $\boldsymbol{\rho}_1$ evolves in time it would move through all possible experiential states resulting from the superposition of the experience eigenstates with different values of the coefficients $a_i$. The Universal thus comprehends within itself all possible states of experience of all individuals at all times.

Therefore, we can say that the individual is a particular sub-stream of the Universal in the sense of being a limitation by projection and not by any actual division into parts. This is quite a novel understanding of the relationship between the individual and the Universal that emerges from our phenomenological approach. Manousakis' Universal therefore is more the collective than the Universal. But, in the long run, since all collectives also finally find their place in the Universal in a nested manner (Universal ⊃ collective ⊃ individual), we may say in this sense that Manousakis is correct in referring to the individuals as sub-streams of the Universal. Suffice it to say that the individual, in the course of its identification (or feeling of oneness) with successively larger collectives, has to shed its individuality gradually in the process till it reaches the very limit of such largeness that it identifies itself with all existence, the whole Universe of things and beings, matter and mind, embedded in space-time. This is when the individuality completely drops off (d becomes zero) and the Universal is realized as one's own essential 'being'. Till such time the individuality maintains itself through the various nonvanishing values of the ABCD-observables.

We note that the results in this section would not be affected even if we added the fourth state to the summation in eq. (8) with $a_4(t) \neq 0$ for ephemeral time intervals (much shorter in duration compared to the observation times), so that the dwellings in this state almost go unnoticed and hence the phenomenological normalization etc. can all be done with only the symmetric triplet for the vast number of ordinary individuals characterized by non-vanishing values of the individualizing observables. Significant departures in the weight factor would occur only if a significant number of individuals spend a considerable amount of time (comparable to the dwell time in any other state) in the fourth state for reasons to be discussed below.

## 7. Interpreting the fourth state

We now come to the most important part of our analysis of experience basing on spin-like observables, which was undertaken to see whether we can get any new understanding or interpretation of Schrödinger's endorsing remark [19] regarding the possibility of the individual experiencing the Universal Being as it is, by becoming identical with it.

First of all, the antisymmetry of the fourth state tells us that it is the only state possible with complete symmetry between the subject and the object in all other respects since the full fermionic state should be anti-symmetric with existence treated as a fermionic quantum number. The triplet becomes possible only because of distinguishability of the subject and the object on the basis of another quantum number, namely, consciousness of existence. Thus it is a state of complete identicality of the existence aspects of the subject and the object. The subject does not know itself to be different from the object.

Secondly, because all individualizing quantum numbers $I^n$ of the nth individual in this state become zero, He is no longer an individual. From eq. (6), His state of experience is given by:

$$|\omega_4^n> = |\Omega_4; I^n = 0> = |\Omega_4; A_\pm^n{=}0, B^n{=}0, C^n{=}0, D^n{=}0> \quad \ldots \ldots \ldots (14)$$

Now, the fourth states of the individual and the Universal experience arising out of interaction between the individual or Universal subject and the corresponding object are given respectively by

$$|\omega_4\rangle = |0, 0\rangle_i = (1/\sqrt{2})\{|+½, -½\rangle_i - |-½, +½\rangle_i\}$$

and

$$|\Omega_4\rangle = |0, 0\rangle_U = (1/\sqrt{2})\{|+½, -½\rangle_U - |-½, +½\rangle_U\}$$

where, the subscripts i and U are introduced for distinguishing the states. Since there is a postulated one-to-one correspondence between the individual and the Universal experiences, there must be the Universal Subject-Object interaction (equivalent to eq. (4) for the individual case), viz.

$$V_U = K_U\, \mathbf{E_s} \cdot \mathbf{E_o} \qquad \ldots \ldots \ldots (15)$$

to account for the similar experiences. The **consciousness of Universal existence** characterizing the Universal Subject (having existence quantum number $E_S = ½$) is what distinguishes It from the universal Object (having existence quantum number $E_O = ½$) in the Universal triplet $|\Omega_i\rangle$, i=1, 2, 3, while in $|\Omega_4\rangle$, the anti-symmetry of the full Universal state demands complete symmetry or interchangeability in all other respects between them.

The fourth state being one with the Universal fourth state is independent of the individual's experiences and is therefore an ever-present state of consciousness as the unchanging and unchangeable background of all the experiences in the other three states and therefore may serve as an alternative route for all the allowed transitions amongst the states as remarked earlier. The only thing to be borne in mind is that the transit through the fourth state occurs extremely fast bordering almost on non-recordability and thus is seldom registered. However, if one wishes to experiment, one may do so oneself by repeated practice of remaining alert and aware till the very last point of entering into Dream from waking and vice versa. The truth of the transit through this fourth state ($|\omega_1\rangle \to |\omega_4\rangle \to |\omega_2\rangle$ and vice versa) can only be verified by one's own prolonged and assiduous practice of such awareness. It is difficult for most of

us even to re-enter a particular dream experience willfully, what to speak of waking up on the verge of deep sleep!

In any case, If we can have neural probes sensitive enough to register a temporary cessation of all thoughts (i.e. the state of thoughtlessness or pure consciousness) for an extremely fleeting interval, then also we may get the verification of the above fact. However, our current low-frequency brain wave probes have been able to register cognitive time scales upto about 300 msec[35, 36] for objective experience in the waking state and thus if the dwell time in the fourth state is less, it remains objectively unobservable.

However, subjective experiences may very well go all the way down to about a few Hz in the Deep Sleep state. Ideally, in the fourth state of thoughtless consciousness the frequency of neuronal oscillations should vanish and therefore the EEG should ideally become flat corresponding to a 'timeless experience'. But in reality, the very application of the probes will undoubtedly lead to some very feeble objective awareness state thereby registering some ultra low frequency oscillations corresponding to that and thus will deprive ourselves of making any objective experimentation on Consciousness.

It seems that at some point in our investigation, we have got to shed the objective approach to reality and make a smooth transition to the subjective approach at the borderline between intellect and intuition, if we are to 'understand' or experience consciousness *per se*. Since understanding anyone else's consciousness does not give one much benefit in regard to one's own beyond a certain elementary level of similarity. These are all facts which we have to grapple with one day or the other individually as well as collectively for progress of our understanding. The truly scientific approach as an impartial investigation of the nature of Reality would be to have an open mind with regard to inputs from all fields of research including the so called inexact sciences like sociology and psychology and philosophy and then take a course of action as would unify all aspects of experience.

A relook at the parallelism between the individual and the universal (eq. (4) and (13) and the states $|\omega_4\rangle$ and $|\Omega_4\rangle$) tells us that not only does the individual become one with the Universal in the fourth state by shedding all individuality, the Universal also in its turn, sheds all Universality and becomes one with the Absolute Being that is beyond any description. The Individual thus becomes identical with the Absolute which simultaneously comprehends all but is comprehensible by none, because none else is there to comprehend it as an object of comprehension. This occurs due to the vanishing of all quantum numbers which characterize the individual or the Universal experiences. The individual becomes indistinguishable from the Universal and the Universal becomes one with the Supreme Absolute that baffles all attempts at description.

We may note here that the main difference between Deep Sleep and the fourth state is in the reversal of the complete ignorance in the former($c=1$) to the complete knowledge($c=0$) in the latter. This may be interpreted to be due to the EPR-like correlatedness of the subject and the object in the fourth state wherein knowledge of one leads to the knowledge of the other. Therefore, even though $|\omega_3\rangle$ itself is a maximally entangled state, the ignorance makes perception impossible and thus one neither knows oneself nor any object. The same argument also holds for the other two states of the Bell basis formed by superposing $|\omega_1\rangle$ and $|\omega_2\rangle$. This means that in the fourth state if the subject knows itself completely, then it knows the object also completely. The subjective awareness is correlated with (awareness of) objective existence. These remarks on $|\omega_4\rangle$ apply equally well to $|\Omega_4\rangle$ also in relation to $|\Omega_3\rangle$.

Thus, in this fourth state, Existence becomes one with Consciousness and the only Experience that we may perhaps speak of is one of Bliss of Existence-Consciousness, since it is only in states of experience of extreme joy that the individual experiences complete self-forgetfulness or self-absorption in the experience of bliss– as happens, for example, in the orgasmic experience. Momentarily though, in the height of orgasmic bliss one does indeed becomes united with bliss itself forgetting all individuality. Thus we can say that in such states all individualizing quantum numbers momentarily assume the zero value, since one knows nothing else but bliss alone. It is a momentary state of union of consciousness with bliss or experience of consciousness itself as bliss. Similar is the experience in the fourth state when we consciously apply certain spiritual techniques like deep meditation and absorption to practice the gradual reduction of the individualizing quantum numbers to zero value and to become established in a state of 'thoughtless consciousness' wherein the mind as a bundle of thoughts is completely annihilated. When one succeeds in remaining absorbed in this fourth state for a longer period, there results a united experience of 'Existence-Consciousness-Bliss Absolute' – Absolute, because it is experienced as the One undivided Whole, which is simultaneously the mode of experience (i.e. Existence), the experiencer(i.e. Consciousness) and the experienced (i.e. Bliss).

## 8. Discussion and conclusion

In this work, we have successfully represented all individual experiential states in terms of eigenstates of a pair of interacting spin-like observables. The interpretations given here not only bring out the kind of psychophysical parallelism envisaged by Pauli and Jung, but also at the same time, bring to close focus the quintessential *Upanishadic* thoughts so much lauded by Schrödinger, Schopenhauer and others who have gone deep into their significance. The present study therefore, may act as a bridge between Quantum Theory and Philosophy proper. The present work is a positive step forward in the direction of finding a true unification of all knowledge at the 'source level' since it deals with the issues of experience in the broadest possible terms by treating subject-object duality itself using quantum mechanics.

In summary, our main postulates in this work have been:

a) All individual experience results from the interaction between the subject and the object.

b) The subject and the object both have two primary states of projection– namely, the external or physical and the internal or mental, and therefore, can be treated as quantum mechanical two-state (spin- ½) systems.

c) The Individual experiences have their source in the Universal and there is one-to-one correspondence between the individual and the Universal states of experience.

d) The individual experiences can be accounted for by the operation of four individualizing observables called the ABCD- observables corresponding to Attitude, Body-identity, Causal Ignorance and Distinctive Ego respectively which take up non-vanishing values. When all of them vanish, the individual becomes indistinguishable from the Universal.

And, the main results have been:

a) The four states of experience emerge from the interaction between the subject and the object– three of them being the triplet of the ordinarily experienced states of waking, Dream and Deep Sleep, while the fourth one is extraordinary, in which all the individual quantum numbers vanish. It is rarely experienced and is the EPR-like singlet state wherein the subject and the object are entangled making knowledge of the one possible from the knowledge of the other.

b) A novel understanding of the Deep Sleep state emerges from the interacting fermion model in which the unconsciousness is seen to be due to lack of contact of the subject with the object because of their being oppositely projected.

c) The individual is identical with the Universal in the fourth state. The Universal in its turn, has all Universal quantum numbers vanishing in the fourth state and therefore is identical with the Absolute which is beyond any description. This establishes the essential identity of the Individual with the Universal.

It reinforces our trust in the versatility of application of the formalism of quantum theory, as it is applied to a domain which was hitherto considered to be exclusively in the realm of psychology and philosophy. The introduction of the individualizing observables is a step forward in the direction of bridging the gap between the exact and the so-called inexact sciences. We remark that the sense in which these observables are used in this essay may not quite tally with the sense in which they are used in the other branches e.g. psychology and management studies etc. because the purpose of introducing them here is purely to generate the individual from the Universal and not just to judge the ability or utility of the individual in a certain situation as required in case of the latter. However, links may be established quite easily between the approaches wherever possible by making appropriate alterations in their definitions in these other branches, since the fundamental ontological character of these observables in our approach makes them more robust.

One important point of departure from traditional psychology is the interpretation of the various phenomena like absentmindedness etc. (hitherto considered as part of the waking experience) as superposition of waking and Dream states. This is but quite natural in the quantum mechanical scheme that we have adopted to describe experience and it aids our understanding of the subconscious mind by bringing it to the forefront of psychoanalysis.

In the process, of course, traditional Quantum mechanics itself suffers a bit, as expected! And, it is in the interpretation of the states that are superposed as simultaneously experienced states, while in the probabilistic interpretation we do not accept simultaneous existence in the superposed states. Instead, we talk of probability of existence or experience of such states. Again, this is not detrimental to Quantum theory in any way. Rather, it may be seen as a real pointer to go beyond the probabilistic interpretation and accept the simultaneous existence of a quantum system in all the superposed states, however absurd it may seem to our classical brain.

Such an interpretation in the case of a free quantum object has been proposed recently[37] where it is shown that the probabilistic interpretation keeps intact our classical notion of a point particle through the introduction of probabilities, but it is plagued with illogical and unsatisfactory features. The most glaring of them is the fact that individual tachyonic de Broglie waves (which are branded unphysical) are superposed to get a bradyonic wave packet which represents the physical particle! It is argued that the free quantum object must be interpreted to have a pervasive existence prior to any interaction or measurements. Granting the quantum system a simultaneous existence (not just a probability of existence) in all the available states would pave the way for clearing up all the mess regarding non-locality and quantum entanglement.

The EPR-correlated fourth state of the subject-object combine may be taken to be the starting point of a consciousness-based cosmology which will contain all the currently acceptable cosmologies as special cases. Cosmology must have the subject built into its structure from the very beginning alongside the object (if not prior to it!) in view of its primal nature as shown in the present essay, because the subject and the object form the dual aspects of all experience.

The relationship with the many worlds/minds interpretation of quantum theory and also the building up of quantum theory in the lines indicated by Manousakis are other aspects which may be worked out keeping in view the general framework introduced in this work.

Possible future explorations may be made by relaxing the orthonormality condition eq. (13) between individuals to account for the more occult-like psychological phenomena such as 'simultaneous perception', 'thought transference', 'deputy perception' etc. discussed by Atmanspacher *et al* [16]. Explanations may also be given for the phenomena like 'metempsychosis' which may finally find their rightful place as fields of scientific investigation on the basis of the robustness of the individuality or the distinctive ego, which is destroyed only in its final dissolution in the Universal or the Absolute. But, at this stage, these are more of a speculation, although, the seeds of their being understood quantum mechanically are very much contained in the analogical framework proposed here.

## 9. Acknowledgements

The author wishes to thank L. P. Singh for many useful discussions and remarks on the manuscript. Thanks are also due to Jagannath Saha, who has been an early inspiration for the author to take up deep study of the philosophical underpinnings of Quantum Theory.

Press, London, 1953).

Please note that the term subconscious used here includes and goes beyond the preconscious of Freud and it generally denotes the entire field of experience between the fully conscious and the fully unconscious. The preconscious is the covering lid of the subconscious and the rest of the subconscious is what Freud took to be the unconscious which is also a little bit different from the sense in which the term unconscious is used here.

[28] A. L. Edwards, '*Techniques of attitude Scale Construction*', Appleton Century-Crofts, New York, (1957);

[29] M. E. Shaw and J. M. Wright, ' *Scales for measurement of Attitudes*', (McGraw- Hill, 1967).

[30] J. McIver, and E. carmines, 'Unidimensional scaling', (SAGE, 1981).

[31] R. Gordon, '*Unidimensional Scaling of social variables: Concepts and Procedures*', (Free Press, new York, 1977).

[32] J. G. Snider and C. E. Osgood, '*Semantic Differential technique: A Sourcebook*',Aldine, Chicago,(1969);

[33] S. Himmelfarb, 'The Measuremen of Attitudes' in "*psychology of Attitudes*", Eds. A. H. eagly and S. Chaiken, (Thomson/Wadsworth, Pp. 23-88, (1993).

[34] This would lead to a density matrix more like the standard quantum Mechanical one with different individuals occupying the same state with intensities proportional to the $|\text{amplitude}|^2$. The overlap of the states of diferent individuals implies that they some of the individualising observables have their values matching each other.

[35] C. Basar-Eroglu, D. Strüber, M. Stadler and E. Kruse, ' Multistable Visual perception induces a slow positive EEG wave', International Journal of Neuroscience, 73, Pp 139-151, (1993).